\documentclass[12pt]{article}

\usepackage{amsmath}
\usepackage{graphicx}

\alph{footnote}

\author{Erik Volz\footnote{Department of Sociology, Cornell University, email:\emph{emv7@cornell.edu}}}
\date{\today}

\title{Tomography of random social networks\footnote{The Cornell email network was provided by Cornell Information Technologies (CIT). Special thanks to Jim Howell and Don Macleod at CIT for help preparing the data. Thanks to Matt Salganik, Douglas Heckathorn, and Stephen Strogatz for valuable comments.}
}

\begin{document}
\maketitle 

\begin{abstract}
We study the statistical properties of large random networks with specified degree distributions.  New techniques are presented for analyzing the structure of social networks. Specifically, we address the question of how many nodes exist at a distance from a given node. We also explore the degree distribution of for nodes at some distance from a given node. Implications for network sampling and diffusion on social networks are described.  
\end{abstract}

\section{Introduction}
 
 Random network models have a long history in the social networks literature. Rapoport et. al. were the first to propose random graphs as models of social networks~\cite{rapo1,rapo2,rapo3}, while simultaneously the basic theory of random graphs was established in the mathematics literature by Erd\H{o}s et. al~\cite{erdo1}. Thereafter, periodic efforts were made to specify with greater detail the random or statistical nature of social networks, for example with the \emph{biased random net} theory of Frank~\cite{franStra1}, Skvoretz~\cite{skvo1},  Fararo~\cite{fara1,fara2}, and others.
 
 More recently, significant contributions have been made by statistical physicists, especially regarding the aggregate statistical attributes of networks~\cite{newmWattStro2,pastRubiDiaz1, newm3}. The degree distribution has been shown to be one of the most important features of a network in determining network structure. Consequently, random networks with specified degree distributions have been proposed as a model of large, complex social networks~\cite{newmWattStro1,holmEdliLilj1, newm1, newmPark1}.
 
 In this article, we describe techniques for revealing subtle aspects of network structure, taking as given a certain degree distribution. Our method relies on \emph{network tomography}~\cite{kaliCoheBenaHavl1}, the idea of mapping out a network layer by layer from a single node. The method is described in section~\ref{sec:netwTomo} below. 
 
 The appropriateness of the random graph model must vary from population to population. Certainly a degree distribution does not determine the overall structure of a network. It is possible for a network with a given degree sequence to have extreme differences from a corresponding random network~\cite{newmPark1,volz1,snij1}. But even in such cases, differences are likely to be informative, suggesting unique mechanisms that move a network away from the random regime. 
 
 This work has implications for networks sampling, the study of diffusion and mathematical epidemiology, as well as other dynamic processes on networks. All of these problems involve the marriage of network structure with network dynamics. To answer dynamical questions, it is desirable to specify network structure with greater precision. Unfortunately, even in random networks of the type studied here, namely semi-random networks with given degree distributions, there are many topological questions which remain unanswered. We will focus on two: 1. How many individuals are there at any distance from a given node? 2. Among all nodes at a given distance, what is the degree distribution among those nodes? Example applications are further described in section~\ref{sec:disc}.
 
\section{Network tomography\label{sec:netwTomo}}
 
 In all that follows, we assume a network size \emph{n}, and a degree distribution \emph{$p_{k}$} (The probability of a node being degree $k$ is $p_{k}$). Multiple connections and loops are allowed, however it should be noted that such connections are exceedingly rare for large $n$. Our networks are undirected. Connections within the network are entirely random but for these constraints. 
 
 Having constructed such a network, we can play the following thought experiment. Pick a node, $v_{0}$ uniformly at random within the giant component of the network\footnote{
  A \emph{component} in a network is a maximal set of nodes such that there exists a path between any two of them. A \emph{giant component} is a component which occupies a fraction of the nodes in the network in the limit of large network size. 
 }. We will call $v_{0}$ the \emph{seed}. This node will have a degree $\geq 1$, and a number of neighbors at distance one. Those nodes in turn will have a degree distribution specific to themselves, and a number of connections to other nodes at distance two from $v_{0}$. We can continue in this way, eventually breaking the entire giant component into disjoint sets defined by the distance from our seed. Some nodes may not be enumerated in this way, in which event they fall outside of the giant component. 
 
 What we just described is the basic premise of \emph{network tomography}.Network tomography, originally described in~\cite{kaliCoheBenaHavl1}, is a method for revealing the structure of a random network by exploration, layer by layer, from a single starting node.
 
\begin{figure}
  \begin{center}
   \includegraphics[width = \textwidth]{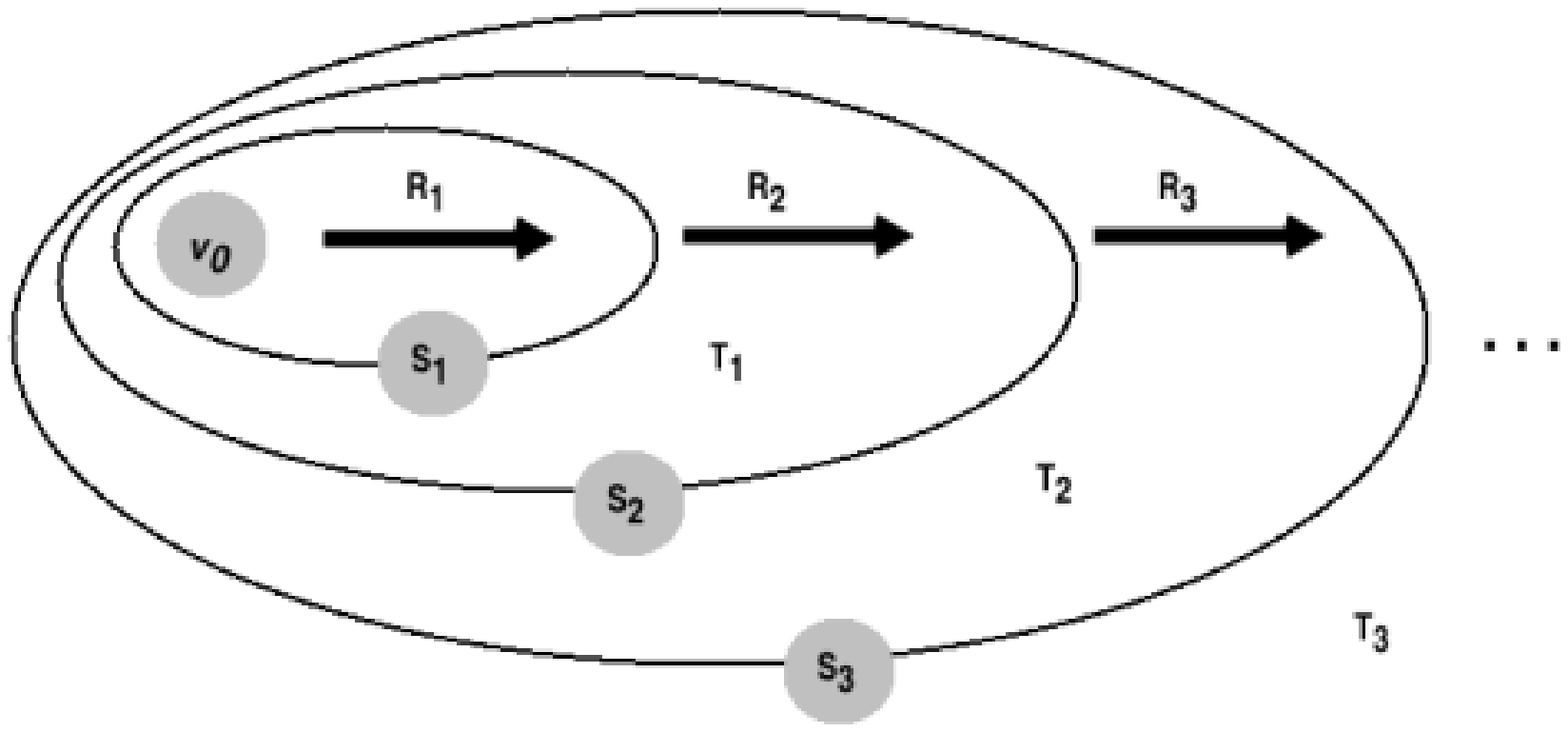}
   \title{Schematic of the network tomograhic method}
   \caption{ \label{fig:schema} This diagram illustrates the tomographic method detailed in the text. Starting from a single node $v_{0}$ we recursively explore nodes at distance $l$ from $v_{0}$. $R_{l}$ is the number of connections going to layer $l$ from layer $l-1$. $S_{l}$ is the number of connections to nodes in layer $l$. $T_{l}$ is the number of connections not connected to nodes in layer $l$ or less. The importance of these quantities is explained in the text.  } 
  \end{center}
 \end{figure}

 Now we can ask a host of questions with consequences for the structure of the network as a whole:
  \begin{itemize}
   \item How many nodes are there at distance $l$ from the seed $v_{0}$?
   \item What is the degree distribution within each layer?
   \item What is the size of the giant component?
   \item What is the degree distribution within the giant component versus outside the giant component?
   \item What is the expected centrality of a seed $v_{0}$ picked at random in this way? What about the centrality of a degree \emph{k} node?
  \end{itemize}
 All of these questions can be answered as outlined below.
 The method is shown schematically in figure~\ref{fig:schema}.
 
 Let $S_{l}$ be the number of connections originating from layer $l$. For example, for $l=0$, $S_{0}$ is the degree of $v_{0}$. Let $R_{l}$ be the number of connections from layer $l-1$ to layer $l$. Finally, let $T_{l}$ be the number of connections originating from nodes \emph{outside} of layers $m\leq l$. 
 
 Let $S_{0} = z_{0}$ where $z_{0}$ is the average degree in the giant component of the network\footnote{
  We can choose any degree for our seed, though some of the statistics we derive will be dependent on this parameter. 
 }. $T_{0} = n z - z_{0} $, where $z$ is the average degree in the network as a whole, and $R_{0} = 0$. To continue mapping out the network, we need a recurrence relation on these quantities: 
 \begin{eqnarray*}
  S_{l+1}  & = & f_{S}(S_{l},T_{l},R_{l}) \\
  T_{l+1}  & = & f_{T}(S_{l},T_{l},R_{l}) \\
  R_{l+1}  & = & f_{R}(S_{l},T_{l},R_{l})
 \end{eqnarray*}

 To proceed further, and determine the exact form of $f_{\cdot}(\cdot)$, we will need to draw on a technique widely employed in the complex networks literature, the \emph{probability generating function}.
  Probability generating functions have found numerous applications to the study of complex networks. The first examples were given in~\cite{newmWattStro2,newmWattStro1}.
 A good general reference to generating function methods is~\cite{wilf1}, and applications of generating functions to branching processes are given in~\cite{harr1} and ~\cite{athrNey1}.

 Probability generating functions are created by transformation of discrete probability distributions into the space of polynomials. We will need just one generating function corresponding to our degree distribution:
 \begin{equation}
  g(x) = p_{0} + p_{1} x + p_{2} x^{2} + \cdots
 \end{equation}
 Frequently we find that generating functions converge to simple algebraic functions, in which cases we can perform any operation on the algebraic version of the generating function instead of the series expansion. This constitutes one of the primary uses of probability generating functions.
 
 In the examples that follow we will concern ourselves with two easy to study degree distributions:

  1. Poisson. This is the degree distribution of classical random graphs as studied by the Erd\H{o}s and Rapoport among others. $p_{k} = \frac{z^{k} e^{-z}}{k\!}$. This is generated by 
  \begin{equation} 
   g(x) = e^{z (x - 1)}  \label{eqn:poisson} 
  \end{equation}
  
  2. Exponential. $p_{k} = (1 - e^{-1/z}) e^{-k/z}$. This is generated by 
  \begin{equation}
   g(x) = \frac{1 - e^{-1/z}}{1 - x e^{-1/z}} \label{eqn:exponential} 
  \end{equation}
  See~\cite{newm3} for a derivation of these generating functions.

 Returning to the tomographics problem, consider the probability that a connection emerging from layer $l$ will go to a node in layer $l+1$, given that the connection does not go to layer $l-1$. Since our networks are completely random, such a connection has uniform probability of going to any of the ``stubs'' originating from nodes in layers $m>l$, as well as stubs originating from nodes in layer $l$, minus those stubs which are already allotted to layer $l-1$. This gives us the following:
 \begin{displaymath}
  P_{l\rightarrow l+1} = \frac{T_{l}}{T_{l} + S_{l} - R_{l}}
 \end{displaymath}
 
 For convenience, we now define the following quantity:
 \begin{displaymath}
  \alpha_{l} = \alpha_{l-1} \frac{T_{l}}{T_{l} + S_{l} - R_{l}}
 \end{displaymath}
 This is the probability of a conjunction of events, namely that a connection goes to a node outside of layer $l$, given that the connection has not attached to layers $m<l$. 
 
 Note that the probability that a degree $k$ node lies outside the first $l$ layers is the probability that all $k$ of the nodes connections go to other nodes outside of layers $m\leq l$. This is simply $\alpha_{l-1}^{k}$.  
 
 Now it can be asked: What is the average degree of a node outside of layers $m\leq l$? We have
 \begin{equation}
  <k>_{T_{l}} = \sum_{k} \alpha^{k}~ p_{k}~ k /c
 \end{equation}
 where $c$ is the appropriate normalizing constant:
 \begin{displaymath}
  c = \sum_{k} \alpha^{k} p_{k} 
 \end{displaymath}
 The value of our generating function approach is now apparent, as we can easily express the above in terms of our generating function $g(x)$:
 \begin{equation}
 <k>_{T_{l}} = n[ \frac{\rm{d}~ g(\alpha_{l} x)}{\rm{d}x} ]_{x = 1}/g(\alpha_{l}) = \alpha_{l}g'(\alpha_l)/g(\alpha)
 \end{equation}
 
 By similar reasoning, the total number of connections originating from nodes outside of layer $l+1$ is: 
 \begin{equation}
  T_{l+1} = n [ \frac{\rm{d}~ g(\alpha_{l} x)}{\rm{d}x} ]_{x = 1}  = n ~\alpha_{l}~g'(\alpha_{l}) 
   \label{eqn:recRel1}
 \end{equation}
 
 Once this is known, $S$ and $R$ follow easily. $S$ is equivalent to the change in the number of connections between two adjacent layers. $R$ will be the expected number of connections going between two adjacent layers. We have:
 \begin{eqnarray*}
  S_{l+1} & = & T_{l} - T_{l+1}\label{eqn:recRel2} \\
  R_{l+1} & = & S_{l} \frac{T_{l}}{T_{l} + S_{l} - R_{l}} = S_{l} \alpha_{l} / \alpha_{l-1}
 \end{eqnarray*}
 
 This recurrence relation can be solved to any desired depth. Below it will be shown that many interesting quantities can be computed from the sequences of S,T, and R.
        \footnote{
                It is worth noting that the recurrence relation on S,T, and R can be simplified to
  a recurrence relation on just two variables, due to that S is not a function of itself. Specifically, by eliminating S, we get
                \begin{eqnarray*}
                T_{l+1} = n \frac{ \frac{d}{{dx}} [ g(\alpha_{l+1} x) ]_{x = 1}}{g(\alpha_{l})}  \\
                R_{l+1} = \frac{T_{l} (T_{l-1} - T_{l})}{T_{l-1} - R_{l}}
                \end{eqnarray*}
                and
                \begin{displaymath}
                \alpha_{l} = \alpha_{l-1} \frac{T_{l-1}}{T_{l-2} - R_{l-1}}
                \end{displaymath}
        }

\subsection{Descriptive statistics \label{sec:descriptiveStatistics}}
 Let's return the questions from section~\ref{sec:netwTomo}. With the simple recurrence relation~\ref{eqn:recRel1} and~\ref{eqn:recRel2} we can now characterize many feature of our network. Once a sequence of values of $S_{l}$, $R_{l}$, and $T_{l}$ have been computed, it is quite simple to determine many things about the structure of our network by plugging in the appropriate values into our generating functions.
 
 Of foremost importance is the size of each layer, that is the number of nodes at some distance from our seed. We know that the probability of a degree $k$ node being outside layer $l$ is $\alpha_{l}^{k}$. Then the probability of a degree k node being within layer $l$ is $\alpha_{l-1}^{k} - \alpha_{l}^{k}$. So, choosing a node at random, the probability of that node being in layer $l$ will be $\sum_{k} p_{k} (\alpha_{l-1}^{k} - \alpha_{l}^{k})$. Translating this into our generating function language, and multiplying by the population size $n$, we have
 \begin{equation}
  n_{l} = n (g(\alpha_{l-1}) - g(\alpha_{l})) \label{eqn:stratSize} 
 \end{equation}
 
 The size of the giant component is even easier to derive. Let $\alpha_{\infty} = \lim_{l\rightarrow \infty} \alpha_{l}$
 \footnote{
  It is interesting to note that $\alpha_{\infty}$ corresponds to the probability of a connection not being to the giant component, u, as derived by Newman et al. in~\cite{newmWattStro2}. The way that this quantity is computed is somewhat different.
 }. This is the probability that a connection goes to a node at distance infinity from the seed, or in other words is outside of the giant component. The probability that a degree k node is outside the giant component is then $\alpha_{\infty}^{k}$. Following similar reasoning as above we find the size of the giant component to be 
 \begin{equation}
  n_{gc} = n (1 - g(\alpha_{\infty}))
 \end{equation}

 As we move outward from our seed, we find that the degree distribution changes within each layer of the network. Initially the average degree tends to increase, as nodes are connected to with probability proportional to degree. But quickly high degree nodes are exhausted, and the average degree within a layer decreases sharply.
 
 In the l'th layer the probability of a node being degree k given by
 \begin{eqnarray}
  p_{k;l} = \frac{p_{k}}{c} (\alpha_{l-1}^{k} - \alpha_{l}^{k})\\
  = \frac{p_{k}}{c} \Big( 1 - \frac{T_{l}}{T_{l} + S_{l} - R_{l}} \Big) \alpha_{l-1}^{k} \\
  = \frac{p_{k}}{c} \frac{S_{l} - R_{l}}{T_{l} + S_{l} - R_{l}}~  \alpha_{l-1}^{k} \label{eqn:degDist}
 \end{eqnarray}
 where $c$ is the appropriate normalizing constant for the degree distribution. When $\alpha$ is close to zero, it dominates the above expression, and thus the distribution converges to a power law as we move away from the seed. Of course, if $p_{k}$ decays faster than a power law (e.g. exponentially) then the distribution will theoretically not have the ``fat tails'' characteristic of power-laws for large $k$. This happens regardless of the degree distribution of the network as a whole.
 
 Using identical reasoning as we used to determine the number of nodes in layer $l$, we can determine the generating function for the degree distribution in layer $l$.
 \begin{equation}
  g_{l}(x) = \frac{g(\alpha_{l-1} x) - g(\alpha_{l} x)}{g(\alpha_{l-1}) - g(\alpha_{l})}
 \end{equation}
 Note that $g(\alpha_{l-1}) - g(\alpha_{l})$ is in the denominator to normalize the distribution.

 The degree distribution outside of the giant component is similarly easy to derive:
 \begin{equation}
  g_{gc^{c}}(x) = \frac{g(\alpha_{\infty} x)}{g(\alpha_{\infty})}
 \end{equation}
 And the degree distribution within the giant component is the complement:
 \begin{equation}
  g_{gc}(x) = \frac{g(x) - g(\alpha_{\infty} x)}{1 - g(\alpha_{\infty})}
 \end{equation}
 
 An important sociological consideration is the mean path length and the associated closeness centrality statistic~\cite{wassFaus1,scot1}. Having chosen a seed, we can compute the average distance to other nodes in the network using the quantities calculated above: 
 \begin{equation}
  m_{c} = \sum_{l\geq 1} \frac{l\times n_{l}}{n_{gc}}
 \end{equation}
 This can be considered the expected closeness centrality of a degree $z_{0}$ node in the network, where $z_{0}$ is the degree of our seed. 
 
 %

\section{Theoretical Examples \label{sec:theoreticalExamples}}
 The reader may find it helpful if we illustrate the preceding ideas with a few simple, idealized examples. 
 
 Many social networks fall into one of two regimes. The simplest case is for the degree distribution to be relatively homogeneous, as occurs when individuals connect to one another with uniform probability. This leads to the classical random networks such as those studied by Rapaport and Erd\H{o}s. These are characterized by a symmetric, unimodal distribution, namely the Poisson generated by equation~\ref{eqn:poisson}. In the second regime, we find that a minority of individuals act as ``hubs'' for the network, thereby accounting for the great majority of connections in the network~\cite{bara1}. This leads to highly skewed degree distributions such as power-laws and simple exponentials.
 Although highly idealized, both of these simple cases may have something to teach us about the structure of real social networks. 

 We have explored both Poisson and Exponential networks using simulation and the tomographic methods discussed above. Consider the Poisson degree distribution, with generating function~\ref{eqn:poisson}. Let $n=50000$. 
  
 By combining equations~\ref{eqn:poisson} and~\ref{eqn:degDist} we find that the degree distribution in layer $l$ is generated by
 \begin{eqnarray}
 g_{l}(x) = \frac{e^{z (\alpha_{l-1} x - 1)} - e^{ z (\alpha_{l} x - 1)}}{e^{z (\alpha_{l-1} - 1)} - e^{ z (\alpha_{l} - 1)}} \\
  = e^{z \alpha_{l-1} (x - 1)} \frac{1 - e^{z \alpha_{l-1} x (\gamma_{l} - 1)}}{1 - e^{z \alpha_{l-1} (\gamma_{l} - 1)}} \label{eqn:poissonDegDist}
 \end{eqnarray}
 where $\gamma_{l} = T_{l}/(T_{l} + S_{l} - R_{l})$.
 
\begin{figure}
  \begin{center}
   \includegraphics[width = \textwidth]{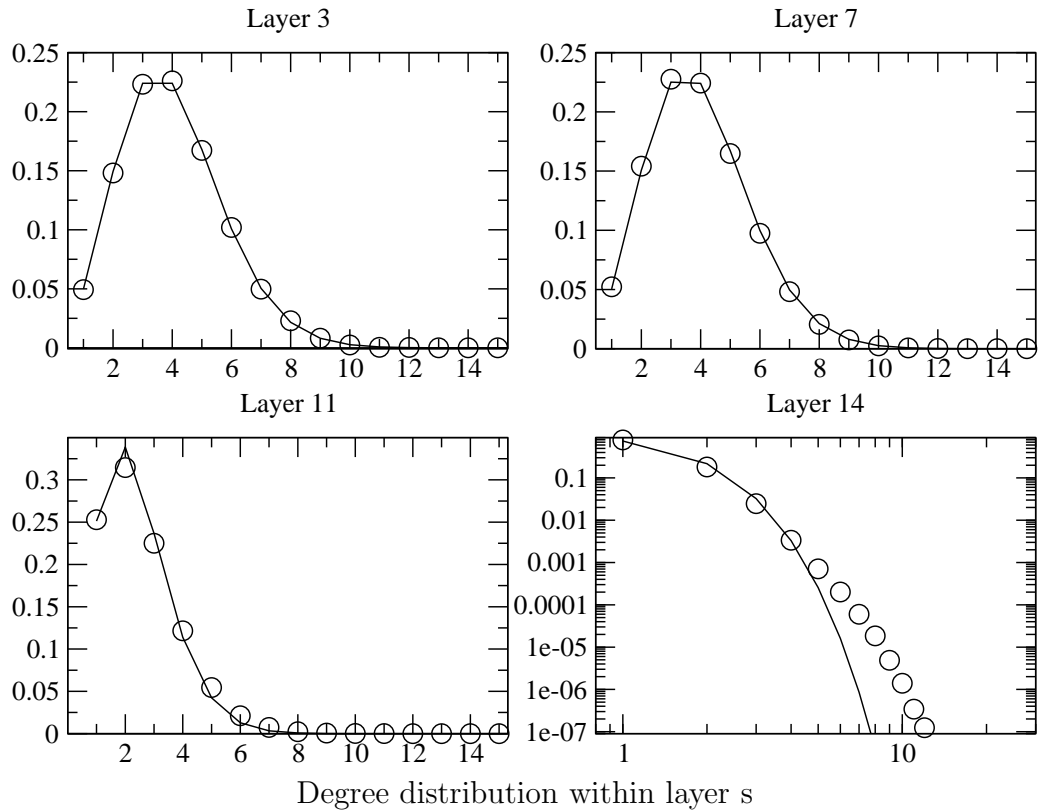}
   \title{Degree distribution within layer s}
   \caption{ \label{fig:poiDD} $n=50000$, Poisson degree distribution, $z=3$. Data points are the average of 40 generated networks with 20 trials per network. Solid lines represent the theoretical prediction given by~\ref{eqn:poissonDegDist}. } 
  \end{center}
 \end{figure}

 It can be verified that this satisfies the requirements for a probability generating function, namely that it has a series expansion, and that $g_{l}(1) = 1$. 
 Figure~\ref{fig:poiDD} shows the degree distribution for $z=3$ at various layers. The solid lines represent the theoretical solutions given by~\ref{eqn:degDist}, and the points, where present, mark the results of simulation. 40 networks of size $n=50000$ and with Poisson degree distribution, $z=3$ were generated. For each network 20 seeds were chosen independently, and the network was mapped out from each. Averaging these simulations yield the data points shown. 

\begin{figure}
  \begin{center}
   \includegraphics[width = \textwidth]{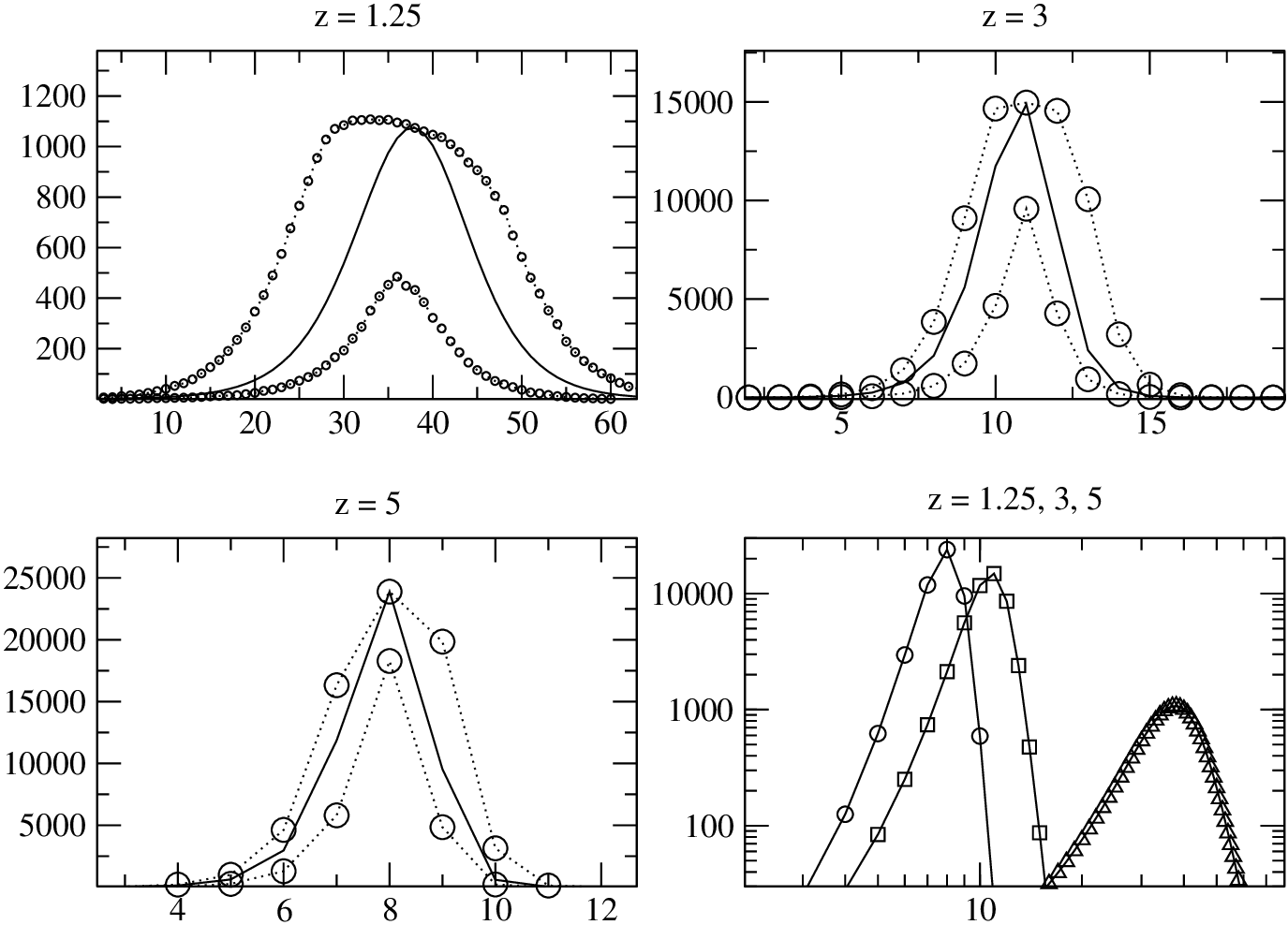}
   \title{$n_{l}$, the number of nodes within each layer}
   \caption{ \label{fig:poiStrat} $n=50000$, Poisson degree distribution, $z=1.25,3,5$. Data points show the 10th and 90th percentile for 40 randomly generated networks with 20 trials per network. Solid lines represent the theoretical prediction given by~\ref{eqn:poiStrat}. }
  \end{center}
 \end{figure}

 Furthermore we can explore how the network changes its structure as the mean of the degree distribution, $z$, is swept over a range of values. Figure~\ref{fig:poiStrat} shows the results of one simulation where $z = 1.25,3,5$ and $n=50000$ as before. The average number of nodes at various distances from a randomly chosen seed is shown. Dotted lines represent the results of simulations, while the solid lines represent the theoretical prediction. The dotted line above the theoretical prediction shows the 90'th percentile among simulations. Likewise the dotted line below shows the 10'th percentile. It can be seen that our theory correctly captures the trend as we increase $z$ from 1.25 to 5. 
 
 The theoretical prediction for figure~\ref{fig:poiStrat} is derived by solving our generating function~\ref{eqn:poisson} and using~\ref{eqn:stratSize}. We find: 
 \begin{equation}
  \label{eqn:poiStrat}
  n_{l} = n e^{z (\alpha_{l-1}-1)} (1 - e^{z \alpha_{l-1}(\gamma_{l} - 1)})
 \end{equation}
 where $\gamma_{l} = T_{l}/(T_{l} + S_{l} - R_{l})$. 
 
\begin{figure}
  \begin{center}
   \includegraphics[width = \textwidth]{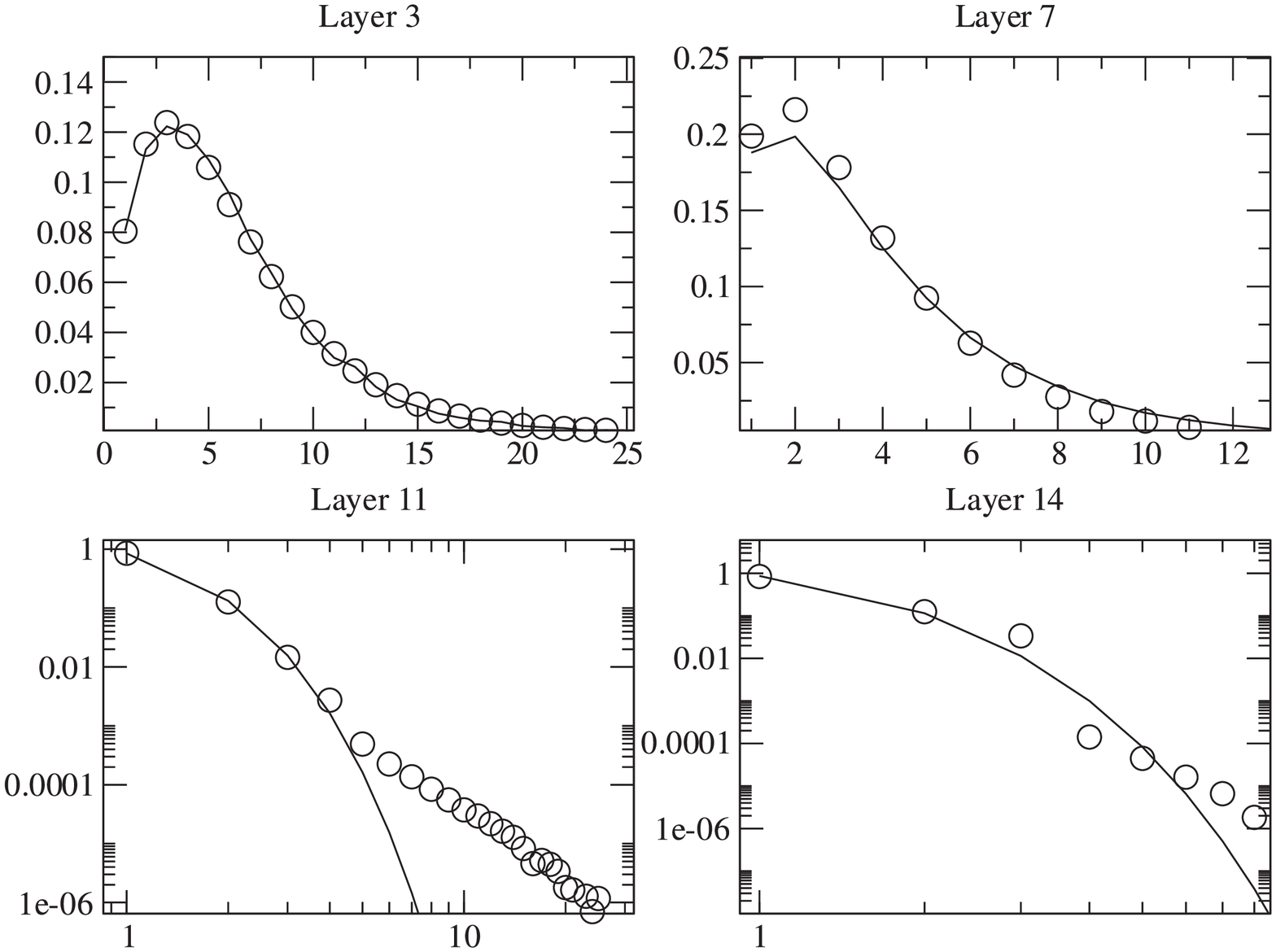}
   \caption{ \label{fig:expDD}$n=50000$, Exponential degree distribution, $z=3$. Data points are the average of 40 generated networks with 20 trials per network. Solid lines represent the theoretical prediction given by~\ref{eqn:degDist}.}
  \end{center}
 \end{figure}

\begin{figure}
  \begin{center}
   \includegraphics[width = \textwidth]{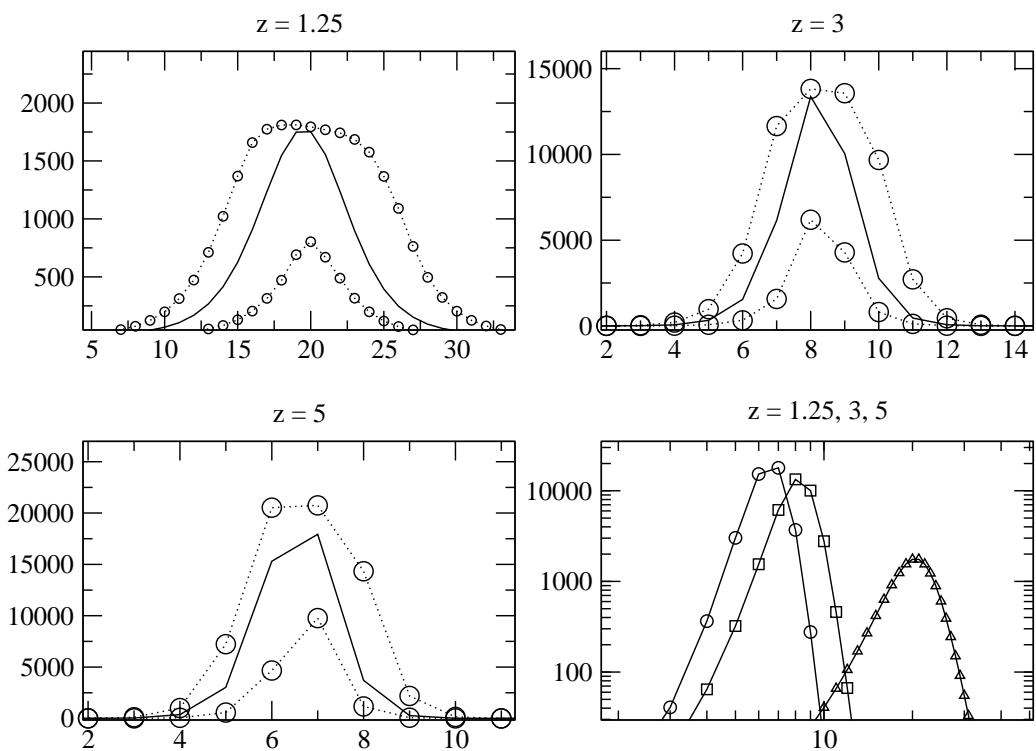}
   \caption{ \label{fig:expStrat}$n=50000$, Exponential degree distribution, $z=1.25,3,5$.  Data points show the 10th and 90th percentile for 40 randomly generated networks with 20 trials per network. Solid lines represent the theoretical prediction given by~\ref{eqn:stratSize}. }
  \end{center}
 \end{figure}

 Figures~\ref{fig:expDD} and~\ref{fig:expStrat} show identical experiments for the exponential degree distribution~\ref{eqn:exponential}. The mathematics is somewhat more tedious for this case, so we omit it here. 
 
 Now viewing the results for the exponential and Poisson experiments, several things bear mention. As we observed above, the degree distribution converges to a skewed exponential or power-law as we move to higher layers in the network. This occurs despite the homogeneous degree distribution of the Poisson networks. In fact, our theory predicts an exponential tail for both of these distributions for high layers. However, we observe the ``fat-tails'' of power laws instead. This is most likely a finite-size effect.
 
 The existence of hubs in the exponential networks lead to several interesting differences with the Poisson networks. It can be seen from the $n_{l}$ experiments that the exponential has a narrower peak than the Poisson. As soon as a path is found from $v_{0}$ to a hub, the rest of the network can be reached in very few steps. It is also interesting that the degree distribution for the exponential random networks has its mode shifted rightward of 0 in the first several layers, thus making its distribution more reminiscent of the Poisson. This is yet another consequence of the existence of hubs in these networks; the higher mode bulge in these distribution represents the existence of higher degree hubs a short distance from $v_{0}$. 
 
\section{Email Network}

\begin{figure}[p]
  \centering
   \includegraphics[width=1.3\textwidth]{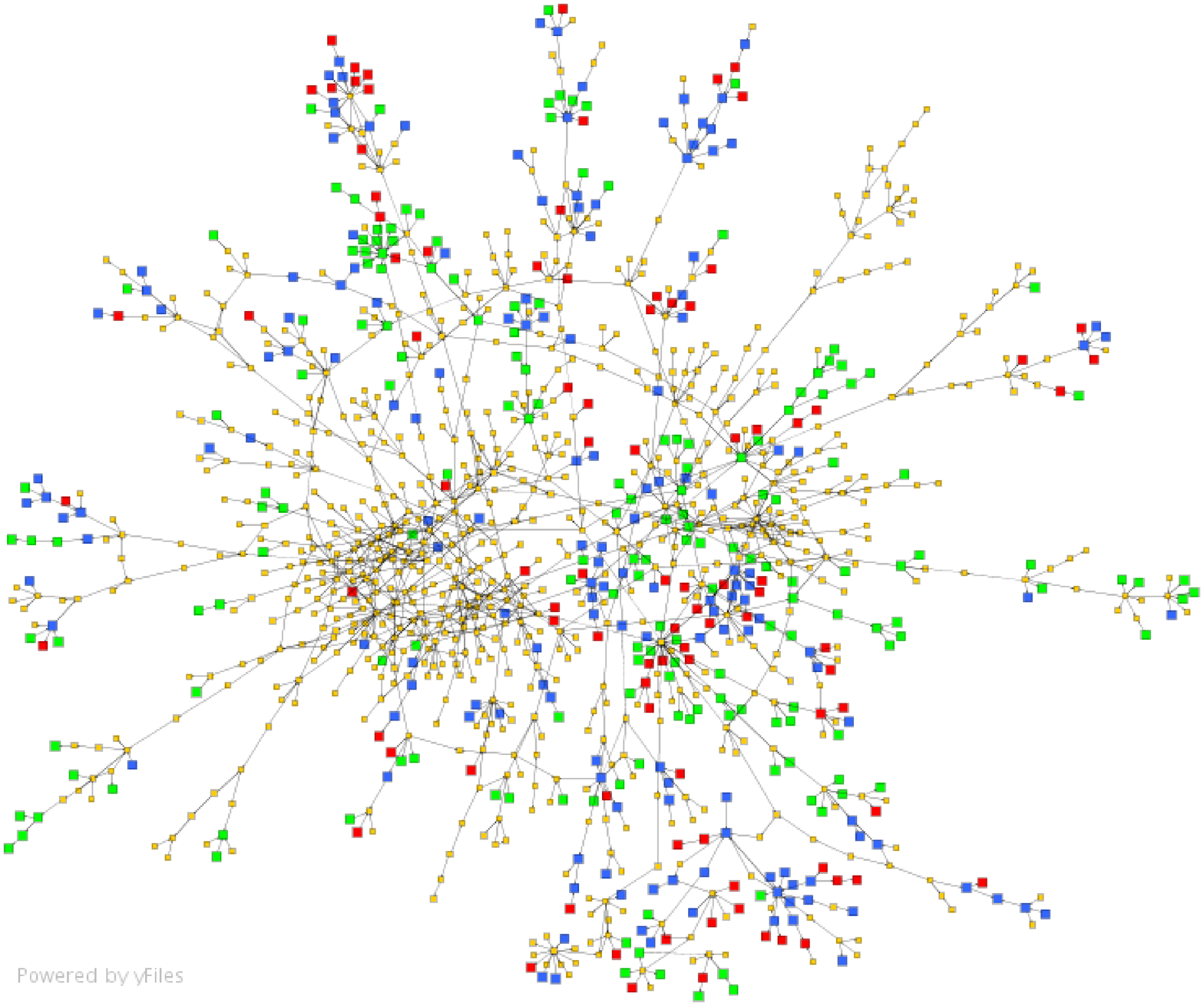} 
   \caption{ \label{fig:emailNet} The giant component from the Cornell email network. Connections in the network represent reciprocal communication within a 24 hour sampling frame. The nodes are color-coded. Blue nodes are faculty, red nodes are graduate students, green nodes are undergraduates, and yellow nodes are everyone else, mainly administrators. The network 2607 nodes and 4838 connections. The giant component consists of 1227 nodes.}
 \end{figure}

The ideas presented here can be illustrated with a real social network. The network shown in figure~\ref{fig:emailNet} is the giant component for a one-day sample of email traffic for individuals at Cornell University. This includes a diverse collection of faculty, researchers, students and administrators. The communication linking them is correspondingly diverse, motivated by work, research and social affiliation. 

In communication networks such as these, it is very important to develop a sense of tie-strength between individuals, particularly for email networks, as a great deal of communication does not indicate a meaningful relationship, but merely the spread of cheap information (i.e. ``spam''). Fortunately, there is an easy way to distinguish genuine social affiliation from simple information transfer. If persons in the network exchange emails in both directions within the 24 hour sampling frame, that is a strong indication that the conversants are well-acquainted and socially connected. We can then induce a subnetwork by including only those ties which are reciprocal.

 \begin{figure}
  \begin{center}
   \includegraphics[width = .5\textwidth]{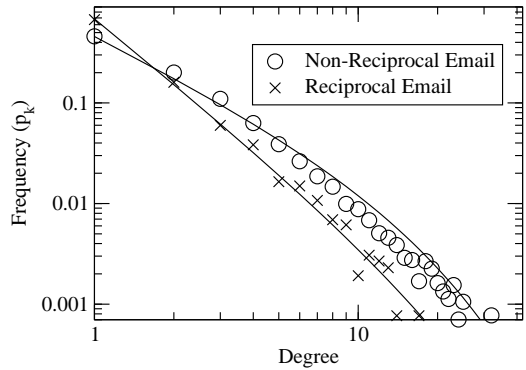}
   \caption{ \label{fig:emailDegDist} Degree distributions for the reciprocal and non-reciprocal email networks. Solid lines show a fit designed to match the average degree of the empirical distribution. The theoretical density is given by equation~(\ref{eqn:powlaw}).}
  \end{center}
 \end{figure}
In what follows, two networks will be considered. The first is the raw communication network, with no distinction made between reciprocal and non-reciprocal communication. For convenience, this will be referred to as the \emph{R/NR} network. This network consists of 14216 nodes with 25040 connections. The giant component of the network occupies 13577 of the nodes~(95.5\%).

The second network consists only of reciprocal email connections and the nodes which have such connections. This will be called the \emph{R} network. This network is much smaller, consisting of only 2607 nodes with 4838 connections. The giant component occupies 1227 nodes~(47.1\%).

The degree distributions for both the R and R/NR networks are shown in figure~\ref{fig:emailDegDist}. Both distributions are evidently power laws, as they lie approximately on a straight line with log/log axes. The solid lines show a fit to these data of a power law density with exponential cutoff:
\begin{equation}
\label{eqn:powlaw}
p_{k} = \frac{k^{-\gamma}e^{-k/\kappa}}{Li_{\gamma}(e^{-1/\kappa})}, k\geq 1
\end{equation}
where $Li_{n}(x)$ is the nth polylogarithm of x. To apply the tomographic theory, we need the generating function for this density. This is given by
\begin{equation}
 g(x) = Li_\gamma(x e^{-1/\kappa}) / Li_\gamma(e^{-1/\kappa}).  \label{eqn:powerlaw}
\end{equation}
When applying the tomographic theory, it is possible to use the empirical degree distribution, but as the theoretical distributions appear to fit the empirical power laws very well, we will use the theoretical distributions instead. 
 \begin{figure}
  \begin{center}
   \includegraphics[width = .5\textwidth]{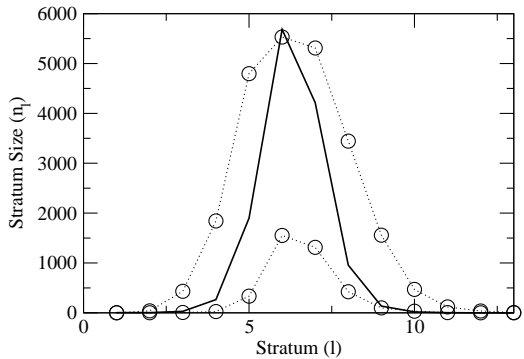}
   \caption{ \label{fig:nonRecipStrat} Theoretical (solid line) and empirical (dotted line) stratum sizes for the R/NR email network. This network includes both reciprocal and non-reciprocal communication within the 24 hour sampling frame. The upper dotted line represents 90th percentile stratum sizes picking a \emph{seed} from the network uniformly at random. The lower dotted line represents the 10th percentile. }
  \end{center}
 \end{figure} 
Figure~\ref{fig:nonRecipStrat} shows the stratum sizes predicted for the R/NR network using equation~(\ref{eqn:stratSize}) (solid line). The dotted lines above and below the theoretical prediction are the actual 90th and 10th percentile stratum sizes from the R/NR network. The theory matches observations fairly well for the R/NR network.
 \begin{figure}
  \begin{center}
   \includegraphics[width = .5\textwidth]{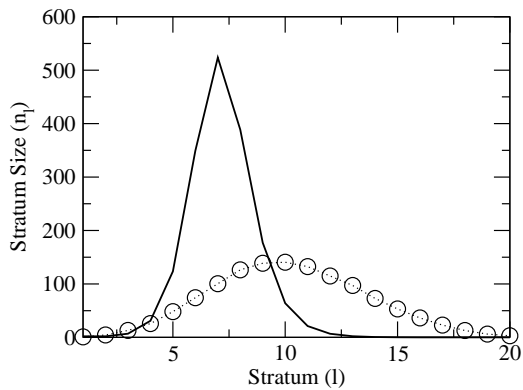}
   \caption{ \label{fig:recipStrat} Theoretical (solid line) and empirical (dotted line) stratum sizes for the R email network. This network includes only reciprocal communication within the 24 hour sampling frame. The dotted line represents the mean empirical stratum size, selecting a \emph{seed} from the network uniformly at random. }
  \end{center}
 \end{figure}
A very different situation is illustrated by figure~\ref{fig:recipStrat}, which shows the theoretical stratum sizes (solid line) alongside the mean stratum size for the R network (dotted line). There is clearly a great deviation between theory and observation. Nevertheless, this difference is instructive. The R network shows only strong ties, in contrast to the R/NR network which contains both strong and weak ties. Consequently, there are many more social micro-structures in the R network than would be expected in a pure random network. The \emph{clustering coefficient}\footnote{
The \emph{clustering coefficient}, $C$, is defined as the ratio of the number of triads to the number of potential triads in a network: $C={3 \frac{N_{\Delta}}{N_{3}}}$ where $N_{\Delta}$ is the number of triads in the network and $N_{3}$ is the number of connected triples of nodes. Note that in every triad there are three connected triples.
}
, a measure of network transitivity, is much greater for the R network ($C=7.4$\%) than for the R/NR network ($C=1.86$\%). Of course, in a pure random network of these sizes, $C\approx0$. Micro-structures such as these contribute to the deviations seen in figure~\ref{fig:recipStrat} because they push the social network away from the pure random regime on which the network tomographic theory is based. As shown in~\cite{volz1}, clustering has the effect of increasing mean path length and decreasing the giant component size. This is why a more elongated series of stratum sizes is observed in figure~\ref{fig:recipStrat}.    

\section{Discussion \label{sec:disc}}
 The methods discussed here have relevance for disparate areas of networks research. 
 
  Consider the problem of \emph{network sampling}-- the utilization of social networks for surveying a population.  
  Lately methods of chain-referral sampling have been proposed~\cite{dougSalg1, volz2} which model chain-referral samples as random walks on social networks. In general, little is known about the attributes of individuals reached after $n$ steps of such a random walk. Tomographic methods may open a new window on the problem. We can now compute the expected properties of a node at a given distance from our starting point, as well as the probability that a random walk will be at that distance after a given number of steps. This allows us to answer questions such as
  \begin{itemize}
  \item How many different nodes could possibly be reached after n steps?
  \item What is the probability of the n'th node in a chain referral sample having degree k?
  \item What is the probability of being at distance l from our starting point after n steps? 
  \end{itemize}
  It is beyond the scope of this paper to provide answers to these questions, but it is certainly possible using network tomography.
  
  Another potential application is to the study of \emph{network diffusion}-- the study of dynamical processes which spread through a population via network connections. Examples include the adoption of innovations~\cite{vale1,roge1} as well as the spread of information or rumors~\cite{guar1,zane1}. The $\{n_{l}\}$ curves shown above are highly reminiscent of birth and death processes such as the spread of an epidemic through a population of susceptible individuals. In fact, the way we have mapped out our network from  a single node is somewhat like the way an infectious agent may spread through a population from an initial infected.  Previous research~\cite{meyePourNewmSkowBrun1} has investigated the structural properties of diffusion of this sort, e.g. the proportion of the network that is ultimately occupied by infecteds. But it has been difficult to place a timescale on diffusion without resorting to computer simulation. It is hoped that progress will soon be made with the application of network tomography to these and related problems. 
  
  All of these results must be taken with the caveat that real networks may not be organized as simple random networks. As mentioned above, there is no guarantee that a real social network will exhibit the same sequences of $n_{l}$ or $p_{k;l}$ as in the random regime. Extra forces can shape the network topology and push these statistics away from the pure random regime. These statistics can be thought of as something that help characterize the structure of the network, like a fingerprint of its structure. When the statistics deviate from the random regime, it is an indication that unique and potentially interesting forces are affecting the network.
  
  A simple example is furnished by the potential existence of greater than random \emph{transitivity}(i.e. triadic closure), which can certainly affect the number of nodes at a given distance from our seed as well as the degree distribution at that distance~\cite{volz1}. However, with more study it may even be possible to adapt the tomographic method to account for transitivity and other non-random structures within social networks.


\begin{thebibliography}{99}
\bibitem{athrNey1}Athreya, K. B., Ney, P., 1972. Branching Processes. Springer, New York.
\bibitem{bara1} Barabasi,L., 2002. Linked. Perseus, Cambridge.
\bibitem{erdo1} Erd\H{o}s,P.,Renyi,A., 1959. On random graphs. Publicationes Mathematicae 6, 290-297.
\bibitem{fara1} Fararo, T.J., 1981. Biased networks and social structure theorems: part I. Social Networks 3, 137-159. 
\bibitem{fara2} Fararo, T.J., 1983. Biased networks and strength of weak ties. Social Networks 5, 1-11.
\bibitem{franStra1} Frank, O., Strauss, D., 1986. Markov Graphs. Journal of the American statistical association 81, 832-842. 
\bibitem{guar1} Guardiola, X., Diaz-Guilera,A., Perez, C.J., Arenas,A., Llas,M., 2002. Modelling diffusion of innovations in a social network. Phys. Rev. E 66, 026121.
\bibitem{harr1} Harris, T. E., 1963. The Theory of Branching Processes. Springer, Berlin.
\bibitem{dougSalg1} Salganik, M., Heckathorn, D., 2004. Making unbiased estimates from hidden populations using respondent driven sampling. Sociological Methodology (forthcoming)
\bibitem{holmEdliLilj1} Holme, P., Edling,C.R., Liljeros, F., 2004. Structure and time evolution of an Internet dating community.
Social Networks 26, 155-174. 
\bibitem{kaliCoheBenaHavl1} Kalisky, T., Cohen, R., ben-Avraham, D., Havlin, S., 2004. Tomography and stability of complex networks. In: Complex Networks. Springer-Verlag, New York, NY.
\bibitem{meyePourNewmSkowBrun1} Meyers, L.A., Pourbohloul, B., Newman, M. E. J., Skowronski, D. M., Brun-ham, R. C., 2005. Network theory and SARS: Predicting outbreak diversity. J. Theor. Biol. 232, 71-81.
\bibitem{newmWattStro2} Newman,M.E.J., Strogatz,S.H., Watts,D.J., 2001. Random graphs with arbitrary degree distributions and their applications. Phys. Rev. E 64, 026118.
\bibitem{newmWattStro1} Newman,M.E.J., Watts,D.J., Strogatz,S.H., 2002. Random graph models of social networks. Proc. Natl. Acad. Sci. USA 99, 2566-2572.
\bibitem{newm3} Newman,M.E.J., 2003. The Structure and Function of Complex Networks. SIAM Review 45, 167-256.
\bibitem{newm1} Newman,M.E.J., 2003. Ego-centered networks and the ripple effect. Social Networks 25, 83-95.
\bibitem{newmPark1} Newman,M. E. J., Juyong,P., 2003. Why social networks are different from other types of networks. Phys. Rev. E 68, 036122.
\bibitem{pastRubiDiaz1} Pastor-Satorras,R., Rubi,M., Diaz-Guilera,A.(eds.), 2003. Statistical mechanics of complex networks. Springer, Berlin.
\bibitem{rapo1} Rapoport, A., Solomonoff, R., 1951. Connectivity of random nets. Bulletin of Mathematical Biophysics 13, 107-117.
\bibitem{rapo2} Rapoport, A., 1963. Mathematical models of social interaction. In: Luce, R.D., Bush, R.R., Galanter, E. (Eds.), Handbook of Mathematical Psychology, vol. 2. Wiley, New York, pp. 493-579.
\bibitem{rapo3} Rapoport, A., 1957. A contribution to the theory of random and biased nets. Bulletin of Mathematical Biophysics 19, 257-271.
\bibitem{roge1} Rogers,E.M., 1983. Diffusion of innovations. FF Shoemaker, New York, 1983. 
\bibitem{scot1} Scott, J., 2000. Social Network Analysis: A Handbook. 2nd ed. Sage, London. 
\bibitem{skvo1} Skvoretz, J., 1990. Biased net theory: Approximations, simulations and observations. 
Social Networks 12, 217-238.
\bibitem{snij1} Snijders, T.A.B., 2003. Accounting for degree distributions in empirical analysis of network dynamics. In: Breiger,R., Carley,K., Pattison, P. (eds.), 2003. Dynamic Social Network Modeling and Analysis: Workshop Summary and Papers, 146-161. National Research Council of the National Academies. The National Academies Press. Washington, DC.  
\bibitem{vale1} Valente, T.W., 1996. Social network thresholds in the diffusion of innovations. Social Networks 18, 69-89.
\bibitem{volz1} Volz, E., 2004. Random networks with tunable degree distribution and clustering, Phys. Rev. E 70, 056115.
\bibitem{volz2} Volz, E., Heckathorn, D., New estimators for chain-referral samples. (under review)
\bibitem{wassFaus1} Wasserman, S., Faust, K., 1994. Social Network Analysis. Cambridge University Press, Cambridge. 
\bibitem{wilf1}H. S. Wilf, Generatingfunctionology, 2d ed. Academic Press, Boston, 1994.
\bibitem{zane1} Zanette, D., Dynamics of rumor-propagation on small-world networks. Phys. Rev. E 65, 041908.
\end{thebibliography}
\end{document}